\documentclass[apj]{emulateapj}
\usepackage{amsmath}
\usepackage{natbib}
\usepackage{graphicx}
\usepackage{dcolumn}

\newcommand{\beq}{\begin{equation}}
\newcommand{\eeq}{\end{equation}}
\newcommand{\bea}{\begin{array}}
\newcommand{\eea}{\end{array}}

\begin{document}

\title{Configurations of Bounded and Free-floating Planets in Very Young Open Clusters}
\author{Hui-Gen Liu$^*$, Hui Zhang, and Ji-Lin Zhou}
\affil{School of Astronomy and Space Science \& Key Laboratory of Modern Astronomy and Astrophysics in Ministry of Education, Nanjing
University, Nanjing 2100093, China, \\
huigen@nju.edu.cn}

\begin{abstract}


{\noindent Open clusters(OCs) are usually young and suitable for studying the formation and evolution of planetary systems. Hitherto, only
four planets have been found with radial velocity measurements in OCs. Meanwhile, a lot of free-floating planets(FFPs) have been
detected. We utilize $N$-body simulations to investigate the evolution and final configurations of multi-planetary systems in very young
open clusters with an age $<10$ Myr. After an evolution of 10 Myr, 61\%-72\% of the planets remain bounded and more than 55\% of the planetary
systems will maintain their initial orbital configurations. For systems with one planet ejected, more than 25\% of them have the
surviving planets in misaligned orbits. In the clusters, the fraction of planetary systems with misalignment is $> 6\%$, and only 1\% have
planets in retrograde orbits. We also obtain a positive correlation between the survival planet number and the distance from the cluster
center $r$: planetary systems with a larger $r$ tend to be more stable. Moreover, stars with a mass $>2.5M_\odot$ are likely unstable
and lose their planets. These results are roughly consistent with current observations. Planetary systems in binaries are less stable
and we achieve a rough criterion: most of the binary systems with $a_{\rm b}(1-e_{\rm b}^2)>$100 AU can keep all the initial planets
survived. Finally, 80\% of the FFPs are ejected out of the clusters, while the rest ($\sim$20\%) still stay in host
clusters and most of them are concentrated in the  center ($<2$ pc).}

\end{abstract}


\section{Introduction}
In current star formation theories, stars initially form in clusters or groups. The same initial mass function (IMF) between field
stars and young embedded clusters provides direct evidence of this\citep{Lada03}. More than 70\% of stars originated from clusters
or groups according to a survey of embedded clusters\citep{Lada03,Lada10}. By reviewing solar system properties, \citet{Adams10}
concluded that our Sun most likely formed in an environment with thousands of stars.

Stars in clusters have basically homogenous parameters (i.e., ages, [Fe/H], etc.); thus searching for planets in clusters,
especially in young open clusters (hereafter YOCs), is very important to understand the formation and evolution of planetary systems.
However, nearly all the detected planets are around field stars, while only four planetary systems are found in open clusters
(hereafter OCs) with radial velocity measurements. Although many groups attempted to find planets by transiting, most of them
had no results (see \citet{Zhou12} for a review and references therein). The four known planets in OCs are: a gas giant planet
around a red giant (TYC 5409-2156-1) in NGC 2423 \citep{Lovis07}, a gas giant planet around a giant star ($\epsilon $ Tauri) in the
Hyades\citep{Sato07}, and two hot Jupiters Pr0201b and Pr0211b in Praesepe\citep{Quinn12}.  The last two are the first hot Jupiters known
in OCs. On the other hand, as compared to bounded planets, several more free-floating planets (hereafter FFPs) are found in OCs.
\citet{Lucas00} detected a population ($\sim 13)$ of FFPs in Orion. \citet{Bihain09} found three additional FFPs in $\sigma $~Orions,
which is a very young OC (VYOC; $\sim 3$ Myr).

Both the detection and non-detection of planets in OCs help us to calculate the occurrence of planets in clusters, which includes
the formation and stabilities of planetary systems. The formation of a planetary system is assured by the IR observation of a
circum stellar disk.  In theory, \citet{Adams06} also show that photoevaporation of proto planetary disks is only important beyond
30 AU due to the median FUV flux of other stars. After planetary systems were formed, star interactions (merges, flybys, etc.),
galactic tides, stellar evolution, inter-planetary interactions, etc., will influence the final orbital architectures of these
planetary systems.

Several previous works investigated the stabilities of planetary systems in clusters. Solving restricted problems,
\citet{Mal11} and \citet{SB01}, and references therein) simulated the influences of assumed flybys on planetary systems, and concluded
that stars passing by with perihelion $\ge 1000$ AU may be negligible, while closer flybys may excite the eccentricities of
planets. \citet{LA98} and \citet{DS01} studied planets in binary systems that encountered stars or other binaries. This revealed
that after the encounter with binary systems, the planetary systems around both single stars or binaries are more easily disrupted
than those after an encounter with single stars. To model the real dynamical environments of clusters, \citet{Spurzem09} used
hybrid Monte Carlo and $N$-body methods to study the evolution of single planetary systems under more realistic flybys from cluster
environments. They found that the liberation rate of planets per crossing time is constant. In their cluster models, a uniform
stellar mass is assumed with no binary included. \citet{Parker12} developed a sub-structured cluster model and simulated the
orbital distributions of single planetary systems in the cluster after 10 Myr. They concluded that, during the dynamical evolution of
YOCs, the planetary systems experienced a relatively violent evolution during the first few megayears, and the fates of these planets depended
strongly on their initial locations, i.e., planets far from the host star can be disrupted easily. Considering the variation of
inclinations in binary systems, \citet{PG09} indicated that about 10\% of the planets in clusters may be affected by the Kozai mechanism.

As mentioned above, most previous works used restricted problems to study the influences of a single flyby event on the planetary
architectures. However, in a real cluster environment, flybys may continuously influence the planetary system. Also the influence of
planetary interactions was ignored due to their single-planet models. As we know, the pumping of eccentricities in closely packed
multi-planetary systems usually leads to dynamical instabilities \citep{TP02,Zhou07}. Planet-planet scattering as well as secular
resonances also influences the orbits of the planets \citep{Nag11,Wu11}. The Kozai mechanism can pump the eccentricities of
planets in binary systems and therefore planetary systems may become unstable due to strong planet-planet interactions
\citep{Mal07a}.

In this paper, we adopt the multi-planetary system models in very young OCs to investigate their different fates as well as the final
orbital architectures of bounded planets. Different from previous works, we use the strict $N$-body simulations to include both
the dynamical evolution of clusters and mutual planetary interactions. We focus on very young OCs with ages less than 10 Myr, so that the dynamical evolution in the cluster is more important than galactic tides or stellar evolution (see also Section 2.1). Multiple flybys are considered in our simulations. We use a more strict model of OCs here than previous works which
contains a mass spectrum and a fraction of binary stars and sets planets located around each star to reveal their stability and
orbital architectures both in binary systems and around single stars. Note that here we first consider the multi-planetary systems in
reasonable cluster environments. Using this model, we intend to investigate how the OC environments mainy influence the
architecture of bounded planets at different locations in the clusters. We will also obtain the fraction of FFPs and their
spatial distribution in OCs.


The structure of this paper is as follows: we introduce our cluster and planetary models in Section 2. In Section 3, we first represent the
dynamical evolution of clusters. After that, Section 4 shows the statistical results of both stars and planets in clusters.  We
study the fates of planets in binaries (Section 5). FFPs in the host cluster and ejected objects are included in Section 6. Finally, we summarize the
main results in Section 7 and discuss some assumptions adopted here.

\section{The Cluster Model and The Initial Setup}
In this section, we represent our VYOC model and the initial setups of planetary systems in clusters.

\subsection{Very Young Open Cluster Model}

The evolution of a general cluster is not only influenced by its internal gravity. Galactic tides and stellar evolutions after the main-sequence phase are still important for the evolution of clusters. The galactic tidal disruption timescale can be evaluated as
$\tau_{\rm tide}=0.077 N^{0.65}/\omega$\citep{GB08}, where $N$ is the number of stars in the cluster and $\omega$ is the angular velocity
around the galaxy center. For a typical cluster with typical $N\sim 1000$ near our solar system, $\tau_{\rm tide}\approx0.3$ Gyr.
Therefore in our model, galactic tides can be ignored in a much shorter timescale of $\sim10$ Myr for VYOCs.


The stellar evolution after the main-sequence phase is also important for the orbital evolution of planetary systems, e.g., the red giant phase
\citep{VL07, VL09}. As we know, the lifetime of stars in the main sequence can be evaluated as a power law by their mass: $\tau_{\rm
MS}\approx10^{10}{\rm yr}(\frac{M}{M_\odot})^{-2.5}$ \citep{Bressan93}, and stars more massive than $16 M_\odot$ would have a lifetime of less than
10 Myr. In the IMF of our cluster model represented next, there are less than 4 stars with a mass larger
than $16 M_\odot$. Due to their large masses, their gravities are important for the cluster but the evolution of these stars is
omitted. We do not consider the residual gas in the cluster due to its limited mass and unknown spatial distribution. The IMF of our
cluster model is taken as two parts \citep{Kroupa02}: \beq N(M)\propto\left\{
\begin{array}{ll}
(M/M_{\odot })^{-1.3}, &    0.1<M/M_{\odot }<0.5, \\
(M/M_{\odot })^{-2.3}, &    0.5<M/M_{\odot }<50.
\end{array}\right.
\label{fmass} \eeq

We truncate the stellar mass $>50M_\odot$  due to the rarity of these stars in the cluster. For example, the most massive star in
Orion is  $ < 50M_\odot$ ($\theta$ Ori C; \citealp{Krause07,Krause09}). Small stars less than $ 0.1M_\odot$ are also ignored in our
model due to their limited gravity, and the very low occurrence of planets around these stars. The IMF of our cluster model is shown
in Figure 1(a) with a total mass of 800-900$M_\odot$.


To make our cluster model more reasonable, we take a fraction of binary systems into account. The fraction of binary systems $f_b$ is
relative to the mass of the primary star. Here we adopted four different fractions of binary systems according to different ranges
of the primary stellar masses\footnote{The references in Equation(\ref{Mfb}) are the following: \cite{FM92}, \cite{Mayor92}, \cite{DM91}, and \cite{Mason09}, respectively.}:
\beq f_b\propto\left\{
\begin{array}{lll}
0.42, &    0.08<M/M_{\odot }\leq0.47, & (\textit{\rm FM92})                                                                                                                                                                                              \\
0.45, &    0.47<M/M_{\odot }\leq0.84, & (\textit{\rm Mayor92})\\
0.57, &    0.84<M/M_{\odot }\leq2.50, & (\textit{\rm DM91})\\
1.00, &    2.50<M/M_{\odot } ,        & (\textit{\rm Mason09})\\
\end{array}\right.
\label{Mfb} \eeq

The separations and eccentricities of the binary systems are set as follows. According to \citet{DM91} and \citet{Rag10}, the periods $P$ (in
days) of the binaries follow a logarithmic Gaussian distribution,
 \beq f(\log_{\rm 10}P)\propto\exp{\frac{-(\log_{\rm 10}P-\mu)^2}{2\sigma}}, \eeq
where $\mu=4.8,\sigma=2.3$. The eccentricities obey a thermal distribution: $f(e)=2e$ \citep{Kroupa08}. Here we only consider 'S' type
planets (planets around each star) in binaries. We constrain the periastrons of binary orbits $\ge 30$ AU, because in these binary
systems, the planets (in orbits $\le 10$ AU) are stable in the restricted three-body problem \citep{MP99}. Meanwhile, 1000 AU is adopted
as the upper limit of the binary semi-major axes. Inclinations are set as 0 for all binaries so their initial orbital planes are all
parallel, while three other orbital elements are chosen randomly. The mass ratio of binary stars is selected as a uniform distribution
according to \citet{DM91}.


In our non-rotating cluster model, each cluster contains 1000 stars in total, located initially in 1 pc$^3$. According to the density
profiles of some YOCs (e.g., NGC2244, 2239, \citealp{Bonatto09}; NGC6611, \citealp{Bonatto06}), the location of these stars can be
described by the two-parameter King model\citep{King66} with the form  $\sigma(r)=\sigma_{\rm bg}+\frac{\sigma_{\rm
0}}{(1+(r/r_c)^2)}$, where $\sigma_{\rm bg}$ and $\sigma_{\rm 0}$ represent the stellar density of the background and in the cluster
center, respectively. $r_c$ is the core radius of the cluster and $r$ represents the distance from the center of the cluster. However, the
King model is a projected two-dimensional density profile. To obtain the three-dimensional symmetric spatial locations of each star, we use a modified Plummer
model here, \beq \rho(r)=\frac{\rho_{\rm 0}}{(1+(r/r_c)^2)^{3/2}}, \label{King} \eeq where $r_c=0.38$ pc in this paper. This is
consistent with the King model after integration in the direction of our sight. The velocities of these stars are set to obey a
Gaussian distribution, with a mean value of  $v$=1 km s$^{-1}$, and a dispersion of $\sigma$=1 km s$^{-1}$. The direction of their velocities is
isotropic; therefore we truncate the distribution where $v<0$, as seen in Figure 1(b). The distribution of stellar location and
velocities, as well as the IMF of stars in our model, corresponds with the normal assumption that the clusters in our model are in
virial equilibrium. So the virial parameter $Q=K/|P|=0.5$, i.e., the absolute value of potential energy $P$, is twice the total
kinetic energy $K$. We set $Q\simeq0.5$ initially via redistributing the velocities of stars.

\subsection{Setups of Planetary Systems}
As mentioned in Section 1, here we focus on OCs with an age less than 10 Myr. These VYOCs are able to give us insight into the
properties of planetary systems around young stars. A large amount of circum stellar disk fraction is found in the VYOCs by $K$-excess observations:
i.e., $30\%-35\%$ of the T-Tauri stars have a disk in the $\sigma$ Ori cluster with an age of $\sim 3 $ Myr\citep{Hernandez07}. Using the {\it Chandra
X-Ray Observatory}, \citet{Wang11} found a $K$-excess disk frequency of $3.8\%\pm0.7\%$ in the 5-10 Myr old cluster Trumpler 15. The
fraction of circum stellar disks limits the formation rate of planets. Combined with stable rate of planetary systems during
subsequent evolution, we can evaluate a planetary system occurrence in these VYOCs.

Due to the large fraction of disks in VYOCs, here we consider planetary systems with two planets around each cluster member.
Considering perturbations from other planets or flyby stars, their orbital parameters can be changed significantly. We use the normal
assumption that all the planets initially formed in circum stellar disks and their angular momentums are always in the same direction
approximately. Here we set the inclination = 1$^\circ$ and eccentricity = 0 for all the planets. The other three angles of orbital
elements are set randomly. We calculate four different initial masses and locations of planets to model different configurations of
planetary systems, as shown in Table \ref{tbl-1}. Planetary systems with two Jupiters represent those with two gas giants in clusters,
called the 2J model.  The one Jupiter and one Earth models with different locations represent more general systems. Note that all
initial planetary systems are stable if they did not experience any close encounter with another star. Hereafter, a planetary system is
unstable when the system loses at least one planet because of close encounters in clusters.

We adopt the MERCURY package for $N$-body simulations\citep{Cham99}. We include the gravities of each star and each planet during
integrations and let the clusters evolve for 10 Myr. During our simulations, we truncate the clusters at 10 pc, i.e., stars and planets
$>10$ pc away from the cluster center are removed as ejected objects. A binary system is thought to be disrupted into two single
stars when its semi major axis is greater than 1000 AU. To judge if the planets are ejected from their host planetary systems, we use
a critical semi major axis of 100 AU, but do not remove the ejected planets. These planets can become FFPs and cruise in the clusters
unless they leave the clusters.


\section{Dynamical Evolution of Clusters}
Before investigating the architectures of planetary systems in the VYOCs, we study the evolution of clusters in this section. Figure 2
shows the variations of the density profiles $\rho$ , half-mass radius $r_h$, virial parameter $Q$, and the percentage of the star
number $N_{\rm S}/N_{\rm Stot}$ with time. Panel (a) gives the densities of stars at 0,1,3,5,10 Myr in the 2J model. Due to the
expansion of the clusters, the density decays with time, which will significantly influence the close encounter rate between planetary
systems and thus results in different dissolution timescales of planetary systems, as shown in Section 3.2. In panel (b), the half-mass radius
$r_h$ (dashed line) increases to 1.5 pc, about three times its initial value. Although the cluster extends to a much larger region in
space, the virial parameter $Q$ (solid line) show it is still at the virial equilibrium ($Q \sim0.5$) at the end of our simulation.
Panel (c) shows that about $3\%$ of the cluster members are ejected out of the cluster in the first 3 Myr in all models. In our models, the
velocity dispersion $\sigma=1\ \rm{km\ s^{-1}}$ $(1\ {\rm km\ s^{-1}}\approx$1\ pc\ Myr$^{-1})$, and about $\sim97\%$ stars with velocity dispersion  $<2\sigma=2\ {\rm km\ s^{-1}}$, adding to the mean velocity $1\ {\rm km\ s^{-1}}$, it will take at least
$ {\rm \sim 3~Myr}$ for a star in 1 pc to arrive at the boundary of the cluster (10 pc as we set in Section 2). Between 3-10 Myr, because of
the dissolution of cluster, $12\%$-$17\%$ of the stars escape from the host cluster and only $80\%$-$85\%$ of the stars still stay in
these clusters after 10 Myr.

To obtain the analytic dissolution timescale, we use the half-mass relaxation timescale(\citealp{Spitzer87}, p. 40): \beq \tau_{\rm
rh}=0.138(\frac{N}{\ln{N}})(\frac{GM_{\rm cl}}{r_{\rm h}^3})^{-1/2}. \label{trh} \eeq
Using thefollowing typical values, $N=1000, M_{\rm cl}\sim800M_{\odot }$,  and $r_{\rm h}=0.9$ pc at $t=3$Myr in Figure 2(b), we obtain $\tau_{\rm
rh}\approx12$ Myr. For isolated clusters here (without the galactic tide), the escape velocity of the cluster $v_{\rm
esp}=2\langle \sigma^2 \rangle{}^{1/2}=2 $km s$^{-1}$,  in the initial Gaussian distribution of velocities, there are
$\sim18.86\%$
of the stars with initial velocities $v\geq v_{\rm esp}$; therefore the dissolution timescale
of the cluster is \beq \tau_{\rm diss}\approx \tau_{\rm rh}/0.1886 \approx 64 {\rm Myr}. \label{tdiss} \eeq
According to this analytic
dissolution timescale, we can estimate that the cluster will lose $7{\rm Myr}/ \tau_{\rm diss} \sim11\%$ of the stars between 3-10 Myr, which is consistent
with our simulation results.


The mass segregation timescale for a star with mass $M$ is also important for the next discussion; here we use a typical expression by Spitzer (1987, p. 74),
\beq \tau_{\rm seg}=(\frac{\bar{M}_{\rm S}}{M})(\frac{N}{5\ln N})(\frac{r_{\rm h}}{\sigma}), \label{tseg} \eeq where $\bar{M}_{\rm S}$
is the mean mass of stars in the cluster. For the typical value used above, we obtain $\tau_{\rm seg}\leq 10$ Myr for an $M\geq2 M_\odot$ star due to energy equipartition in our models.


\section{Planetary Systems around Cluster Members}

\subsection{General Statistical Results}
The general results of the four models are analyzed in detail in this section. Table \ref{tbl-2} represents the number of survival
objects in the clusters in different models. Although 80\%-85\% of the stars are still in clusters, only 66\%-74\% of the planets stay in the
clusters. There are at least 10\% more planets, which were disrupted from their host stars and obtained a large velocity. Finally
these planets escape from the cluster more easily than more massive stars. In these YOCs, we divide the survival planetary systems
into the following three classes.
\begin{itemize}
\item {\it 2pisi systems}. Stable planetary systems maintaining two original planets. 55\%-68\% of the initial systems are in this class.
\item {\it 1pisi systems}. Planetary systems lost one planet and have only one original planet that survived. Only 6\%-18\% of the stars ejected one planet.
\item  {\it pisj systems}. Recaptured planetary systems, which is very rare. In our model, they are less than $1\%$.
\end{itemize}

Hereafter these systems are represented as 2pisi, 1pisi, and pisj systems. Besides the retention of  planetary systems, a fraction of
stars ($\sim$6\%-13\%) lost both planets ($N_{\rm 0ps}$), primarily because of close flybys as near as $<100$ AU. Due to the ejection of planets, only 2.7\%-5\% of the planets are still cruising in the cluster and become FFPs
($N_{\rm FFPi}$), while at least 12\%-21\% of the planets become FFPs outside the host cluster ($N_{\rm FFPo}$). In our model,
$\sim47\%$ of the stars are initially in binary, and after a 10 Myr evolution in the cluster $\sim64\%$ of the binary systems are preserved. After the evolution of the cluster, we also
find a few "naked" stars, $N_{\rm ss}$, without any planetary or stellar companions.

In Table \ref{tbl-2} we see some rough correlation with the bounded energy of planets ($E_{\rm b}$). $N_{\rm P}$, $N_{\rm FFPi}$, $N_{\rm
FFPo}$ and $N_{ss}$ have negative correlations with  $E_{\rm b}$, while $N_{\rm 2pisi}$ is the opposite. The results are reasonable because
larger energy is needed to disrupt planetary systems with higher $E_{\rm b}$. The J5E2 model with a larger $E_{\rm b}$ has 10\%
more survival planets than model J10E4 (also see Figure 6(a)). As \citet{Spurzem09} detailed the influences on single planetary
systems with different semi major axes,  we do not survey the influences of $E_{\rm b}$  in detail here due to our limited four
models. More details about binary systems and FFPs will be discussed in Sections 5.2 and 5.3. Here we focus on the architecture of
survival bounded planetary systems and other properties.

\subsection{Architectures of Planetary Systems}
Figure 3 shows the orbital architectures of planetary systems in different models. In each $a$-$e$ plane, planetary systems are divided into
three classes: 2pisi (green triangles), 1pisi (black circles) and pisj (red squares) systems. The filled symbols represent Earth-like
planets while the open symbols represent Jupiter-like planets. Most planets still stay near their initial locations. The outer Jupiter-like
planets can more easily change their angular momentum than inner planets during flybys, and therefore change their locations or
eccentricities more probably. In panels (b)-(d), Earth-like planets are difficult to eject; therefore most 1pisi systems retain their
filled circles.

The distributions of eccentricities and inclinations of these three classes of planetary systems are shown in Figure 4. Obviously the
2pisj systems have only a negligible fraction to change their initial eccentricities ($\sim$3\%$>0.1$) or
inclinations ($\sim$6\%$>10^\circ$). Meanwhile a large part of the 1pisi systems changed their eccentricities larger than 0.1
($\sim50\%$), or inclinations larger than $10^\circ$($\sim30\%$). Ejected planets become FFPs, which can be recaptured by some other
stars randomly. Hence, the pisj systems tend to have much wider and flatter distributions of both inclinations and eccentricities.

Here we note: for these 1pisi systems, 282 inner planets survived compared with 223 outer ones. The ratio is 1.3:1 on average.
Furthermore, to reveal the different properties of these systems retaining inner or outer planets, we plot Figure 5, adding pisj systems
as blue triangles. Black squares mean 1pisi systems with outer planets that survived, while red circles represent systems where inner planets
survived. In Figure 5(a), we give the final locations of all the systems in the inclination-eccentricity plane, which is divided into four
regions by two lines: eccentricity $=0.1$ and  inclination $=10^\circ$. The red line is inclination $=90^\circ$. Planets above this
red line move in retrograde orbits. Figure 5(b) gives the fraction of systems in the four regions. Nearly all the pisj systems have an
eccentricity $>0.1$, and none of them have a small eccentricity ($<0.1$) and  inclination ($<10^\circ$). In the 1pisi systems where the
outer planet is ejected, about 45\% of the surviving inner planets have small eccentricities and inclinations. However if the inner one is
ejected, only about 25\% of the surviving outer planets have small eccentricities and inclinations, while more than 70\% of the planetary
systems have obviously changed their eccentricities ($>0.1$) or inclinations($>10^\circ$).

The spin-orbit misalignment can be estimated by the Rossiter-McLaughlin effect (see \citealp{Winn10}); thus the inclination study in our
simulations is also interesting. Assuming all stars spin in the same direction invariably for simplification, we have many planets in
misaligned orbits ($>10^\circ$) in clusters. More than $25\%$ of the 1pisi systems that have their outer planets ejected
have the surviving planets in misaligned orbits.  The same misaligned fraction is also obtained in 1pisi systems  that have  inner planets ejected.
  We also find 21 planets ($\sim4\%$) in
retrograde orbits that have surviving single planetary systems. These fractions are quite low except in pisj systems, as shown in the smaller panel in
Figure 5(b). However, it is still higher than the occurrence in 2pisi systems, which contain only 14 ($<0.6\%$)
planets in retrograde orbits. Based on the results here and considering all the planetary systems in the clusters, we calculate the lowest
fraction of planetary systems in VYOCs with a misalignment of at least $\sim 6\%$. Only 1\% have planets in retrograde orbits.




\subsection{$r$-correlations and Mass-correlations}
As pointed out by \citet{Binn87}, the frequency of close encounters is sensitive to the stellar density, which decreases with both
evolution time and the distance from the cluster center $r$ (the same hereafter). As shown in Figure 6(a), the fractions of surviving
planets are very different from that of stars (Figure 2(c)) due to the fast decay of stellar density $\rho$ in the center of the clusters.
In all four models, the fraction of surviving planets decreases in the first 1 Myr. After that, $\rho$ decays quickly and the
decreasing rate becomes smaller and smaller.

Same as the previous time dependence, the stabilities of planets change with different locations in OCs. In the center of the clusters, the
density can be much larger than  in the outer region, which leads to a higher frequency of close encounters (see Equation (\ref{enc})).
Planetary systems near the center of OCs can be disrupted quickly. This is obvious as shown in Figure 6(b). The
number of surviving planets is denoted by $N_{\rm P}$. The distribution of unstable planetary systems with $N_{\rm P}=0$ peaks at 0.958 pc sharply, while a
fatter distribution of systems with $N_{\rm P}=1$ peaks around 1 pc. The peak for stable systems with $N_{\rm P}=2$ is located at 1.29
pc, i.e., these stable systems stay in the outer region compared with the other two unstable systems. This is consistent with the fact that
planetary systems in the inner region of OCs are probably unstable. In the inner 1 pc$^3$, about 40\%, 30\% and 20\% of the planetary systems
have $N_{\rm P}=0, 1$ and 2, respectively. About 80\% (the horizontal dotted line) of the systems with $N_{\rm P}=0, 1, 2$ are concentrated in
2, 3, 4 pc approximately. We call this correlation {\em r-correlations}.

The variations of angular momentum $\Delta L$ for planets in these three systems are also plotted in Figure 7. We can find an obvious
correlation between the maximum $\Delta L$ (with a unit $M_{\rm J}\cdot\sqrt{{\rm GM}_\odot\cdot {\rm AU}}$) and $r$. In the
center of the clusters, more close encounters take more angular momentum away. In Fig.7, we show linear estimations of the upper limit for
\beq|\Delta L|=a(10-r)+b, \eeq
 where the constants $a,b$ for different $N_{\rm P}=2, 1, 0$ are listed in Figure 7. $\Delta L$ of planetary systems in different
three systems at different locations must be less than this limit in our VYOC models.

As shown in Equation (\ref{tseg}), the mass segregation timescale can be less than 10 Myr for stars $>2M_\odot$. The mass segregation leads
to a spatial distribution of stars that correlate to the stellar mass. Massive stars sink into the inner region while small stars
cruise in the outer region. It is similar to the $r$-correlation. However, another influence of stellar mass on planetary stability has the opposite effect.
Large mass stars may hold planets around them more tightly, and thus they need more energy to release these
planets. The stellar mass correlation (called {\em mass-correlations}) combines these two competing effects.

In Figure 8(a) we give the fraction distribution of planetary systems ($f$) for $N_{\rm P}=0, 1, 2$. No planets survived around stars more
massive than $16M_\odot$, because these stars have sunk deep into the center of the cluster in a very short timescale, $\tau_{\rm
seg}<1.2$ Myr, and planetary systems in the center of the cluster are much less stable as pointed out above. The black dashed line columns show
the observational data (the same label as in Figure 6(b)). In order to compare with the
 initial IMF $f_0$ of host stars, we represent a normalized fraction (divided by $f_0$) in Figure 8(b) to highlight the fraction
variation. Systems with $N_{\rm P}=2$ remain stable for stars with $M<2.5M_\odot$. A sharp decrease in the fraction for $N_{\rm P}=2$
and large enhancements for $N_{\rm P}=1,0$ exist at $M=2.5M_\odot$. This critical mass indicates a boundary of about 2.5$M_\odot$ for the
two competing effects. For those more massive stars, most planets around them can still be disrupted due to the heavy density ($\rho$)
in the inner region of the cluster, although they are bounded more tightly. Less massive stars cruise in an
environment with a much lower $\rho$ and can hardly release any planets.  From an observational aspect, three of the four planetary systems in OCs
are found around stars with masses $< 2.5M_\odot$. The left one, $\epsilon$ Tauri, has a stellar mass of $2.7M_\odot$ \citep{Lovis07}. We also predict that more planets ($>80\%$ in our results) will exist  around less
massive stars (0.1-1 $M_\odot$) in these VYOCs.

Although the observational data of planetary systems in OCs are limited, the $r$-correlations and mass-correlations obtained here are
still consistent with observations. In the future, more planetary systems detected in OCs can verify and refine these correlations.


\section{Planets in Binaries}

In our model, an OC contains $\sim47\%$ of the binaries, and the binary fraction decreases with time due to stellar flybys.
Investigating the final fate of planetary systems in these binaries is very helpful for studying the stabilities of planetary systems in the
cluster. In this section, we focus on the orbital variations of binaries (Section 5.1) and especially the planetary systems in binaries
(Section 5.2).

\subsection{Binary Systems}

According to our simulations, a finally binary fraction $f_b$ is achieved
 ($\sim36\%$)
compared with the initial value ($\sim47\%$), and nearly 64\% of the binary systems in clusters remain after 10 Myr. Besides the
binaries ejected outside the clusters, about 15\%-20\%, as shown in Section 3, $\sim$16-21\% of the binaries were disrupted in the clusters due to
close encounters. We estimate the final number of binary systems $N_{\rm b}$ as two parts: the dissolution factor: $\exp(-{\rm
Age}/\tau_{\rm diss})$ and the disrupted factor $\exp(- {\rm Age}/\tau_{\rm disrupt})$, \beq N_{\rm b}=N_{\rm b0}\times\exp(-{\rm
Age}/\tau_{\rm Nb}), \label{Nb} \eeq where Age means the age of the cluster. The timescale for the decay of the binary number can be
calculated as
 \beq \tau_{\rm Nb}=\frac{\tau_{\rm diss}\times\tau_{\rm disrupt}}{\tau_{\rm diss}+\tau_{\rm disrupt}}. \eeq
$\tau_{\rm diss}$ is shown in Equation (\ref{tdiss}). To estimate $\tau_{\rm disrupt}$, we use the encounter timescale obtained by
\citet{Binn87}: \beq \tau_{\rm enc}\simeq 33{\rm Myr}\times(\frac{100{\rm pc}^{-3}}{\rho})(\frac{v}{\rm 1\ km\ s^{-1}})(\frac{10^3 {\rm
AU}}{r_{\rm peri}})(\frac{M_\odot}{m_{\rm t}}). \label{enc} \eeq If the average separation of binary stars is $\bar{a}_{\rm b}$,
assuming the closest distance encountered is $r_{\rm peri}\simeq 2\bar{a}_{\rm b}$, the perturbation of encounters will be the same degree
as the binary companion. These encounters will probably disrupt the binaries. Considering a binary can encounter another single star
or binary, we use 3.5 times the mean mass of star $\bar{M_{\rm S}}$ as the total mass of stars during the encounter $m_{\rm
t}=3.5\bar{M}_{\rm S}$. Taking a typical stellar velocity $v=1\ {\rm km\ s^{-1}}$ , we finally obtain the binary disrupted timescale from
Equation (\ref{enc}): \beq \tau_{\rm disrupt}\simeq 33 {\rm Myr}\times(\frac{100{\rm pc}^{-3}}{\rho})(\frac{500 {\rm AU}}{\bar{a}_{\rm
b}})(\frac{M_\odot}{3.5\bar{M}_{\rm S}}). \label{disrupt} \eeq In our model $\bar{a}_{\rm b}=185$AU, $\bar{M}_{\rm S}=0.8M\odot$. The
stellar density decayed so quickly that here we chose the $\rho\sim60$ pc$^{-3}$ (the value at 1 pc in the cluster with a moderate age of 3
Myr). Substituting these typical values into Equation (\ref{disrupt}), $\tau_{\rm disrupt}\approx53$ Myr. Thus about 17\% of the binaries are
disrupted after 10 Myr, which is consistent with the fraction of the disrupted binary in our simulations.

Based on the observations of YOCs, the binary fraction $f_{\rm b}$ can be estimated by: \beq f_{\rm b}=f_{\rm b0}\times\exp(\rm
-Age/\tau_{\rm disrupt}). \label{fb} \eeq Taking $\tau_{\rm disrupt}=53$ Myr and the initial binary fraction $f_{\rm b0}=47\%$ in our
model, we calculate the binary fraction $f_{\rm b}=39\%$ after 10 Myr according to Equation (\ref{fb}), which is similar to $f_{\rm
b}=36\%$ in our simulations.

The distributions of semi major axes $a_{\rm b}$ and eccentricities $e_{\rm b}$ of survival binary systems are shown in Figure 9. The red
bars show the distribution of the initial fraction($f_{\rm 0}$), while the green bars show the distribution after 10 Myr($f_{\rm 10}$).
The upper panel gives the relative fraction ($f_{\rm 10}/f_{\rm 0}$). The dynamical evolution in clusters can disrupt wider binary
systems more easily due to the lower bounded energy, and the eccentricities of binary systems can also be pumped. Therefore, compared
with the initial distribution, the fraction of binary systems with small $e_{\rm b} (<0.4)$ or large $a_{\rm b} (>200 $AU) decreases
after 10 Myr. Meanwhile, more binary systems with $a_{\rm b}<100$ AU ($>45\%$) or moderate $e_{\rm b} (0.4-0.8)$ ($>55\%$) are left.

Very few new binary systems formed during the evolution of the clusters. In our four models, only 15 new binaries formed in total, with
mean eccentricity$=0.68$ and inclination$=1.66$ rad. Their $a_{\rm b}$ are not regular at all from $60$-$1000$ AU. The contribution to
$f_b$ of these binaries is quite small and can be omitted.

\subsection{Planets around Binary Stars}

During the disruption of binary systems, the planets around each star have different fates. Some collide with host stars, some
are ejected out of the systems, and some still orbit around their host stars. In our models, there are a total of 304 disrupted binary
stars that still stay in clusters after 10 Myr, while others are ejected out of the cluster. Sixty four of them lose all their planets, 59 stars have
only one planet, and the other 181 stars have two bounded planets. We are also concerned with the inclinations of these
planets. For the systems with one planet, 5 in 59 planets are in retrograde orbits. For those with two planets, five systems have at
least one planet in retrograde orbits. In systems containing two planets, planets with mutual inclination $>i_{\rm Kozai}\sim42^\circ$ experience the Kozai effect \citep{Koz62}. There are only six such systems around disrupted stars. Around the original single stars,
only five systems have a mutual inclination of $>i_{\rm Kozai}$. We conclude that, due to perturbation during the disruption of binary
systems, the mutual inclinations of planets around these disrupted single stars are large and have more of a chance of experiencing the Kozai
effect in their long evolution hereafter.

Since lots of the binary systems survived, we next study the planetary systems in these binaries. As we set two planets
around each star, there are a total of four planets in one binary system initially.  Figure 10 shows the left number of planets in binaries
$N_{\rm P}$ in the $a_{\rm b}$-$r$ and  $e_{\rm b}$-$a_{\rm b}$ planes. $r$ is the distance from the barycenter to the center of the
cluster.
As shown in panels (a) and (b), the binary systems with larger angular momentum (larger $a_{\rm b}$ and smaller $e_{\rm b}$) and
smaller stellar density (larger $r$) probably have more survival planets.

Figure 11 shows $N_{\rm P}$ in different $X$-$Y$ planes. We obtain a rough criterion: binary systems with $a_{\rm b}(1-e_{\rm b}^2)>100$AU
seem to restore all four planets, while others more or less lost planets around them. There is still a small fraction for
close binaries that have a small $e_{\rm b}$ and can preserve all the planets very well. As seen in Figure 10(a), the distance from the
center of cluster $r$ has less influence than that in single stars.

Statistically, only 1411 planets (58\% of the initial planets) stay in 604 binary systems after 10 Myr, i.e., each binary star contains 1.17
planets on average. In these 15 newly formed binaries, a total of 22 planes survived; therefore we obtain a much smaller mean planet
number of 0.7 around each star in these newly formed systems. Compared with single stars, which have $\sim1.8$ planets around each
star on average, multi-planetary systems in binary are much less stable than those around single stars.


\section{FFPs and Ejected Objects}

Besides planets bounded around stars, there are fruitful FFPs in our universe \citep{Sumi11}. In our model, there are a few FFPs left
in the cluster; however there is an abundance of FFPs ejected outside the cluster. This is due to the much larger mean velocities of FFPs compared
with the velocities of stars in the cluster; therefore these FFPs are much easier to be ejected. In this section, we will reveal the
distribution of FFPs and other properties of ejected objects.

As the outside Jupiters are much easier to eject out of planetary systems as seen in Figure 3, there are much more Jupiter-like FFPs
(about 2.5 times) than Earth-like FFPs. However, their spatial distributions are similar, as shown in Figure 12, i.e., the CDFs of these
planets with different sizes are nearly the same. Nearly half of the FFPs are concentrated in 2 pc, while 80\% of the FFPs are concentrated in 4 pc. As
shown in Figure 13(a), the fraction of FFPs is relative to their location $r$; we show a fitting curve to model the distribution: \beq
f_{\rm FFPs}=0.003+\frac{0.07}{(r-1.14)^2+0.97}. \label{FFP} \eeq  The maximum fraction peaks at 1.14 pc.

Besides these FFPs cruising in clusters, there are still a large number of objects ($N$=2868 in sum) that were ejected out of
the clusters. The fraction of components of ejected objects is shown in Figure 13(b). 48.2\% of these objects are FFPs ($N$=1381), while
the two-planet systems also have a quite large fraction of 40.6\% ($N$=1164). The very rare one-planet systems are only 2\% ($N$=58). The remaining
9.2\% ($N$=265) are single stars with no planetary or stellar companion around them. Most FFPs are ejected out of the host clusters, as
seen in Table \ref{tbl-2}. $N_{\rm FFPo}$ has a negative correlation to the bounded energy. In all these FFPs, $\sim70\%$ are
contributed by planets in binaries, i.e. planets in binary systems seem to be much less stable.

Based on the above results, the FFPs are likely to be found in the inner region of YOCs, and Jupiter-like FFPs are common. Most FFPs
are ejected outside their host clusters and cruise in the deep universe. For these planetary systems ejected out of host clusters,
more than 80\% keep the same initial number of planets. Few planets ($<10\%$) in these systems have their orbits changed much.
Based on this conclusion, our solar system, which is thought to be formed in a cluster environment, is most likely to form as a similar
current configuration before the ejection from its host cluster.

\section{Discussions and Conclusions}

In VYOCs, in order to repel the galactic tidal force and stellar evolution, we chose an isolated, isotropic, and non-rotating cluster model in
this paper. We investigated the configurations of both multiple and single planetary systems as well as FFPs in VYOCs
without residual gas. In these clusters, a modified King model is adopted to produce the spatial distribution of stars, and virial
equilibrium is satisfied as an assumption. Different from previous works, we add a large binary fraction as well as the IMF of
stars in the cluster model, which is much more realistic than previously adopted models.

Our major conclusions in this paper are listed as follows:
\begin{itemize}
\item After dynamical evolution for 10 Myr, clusters are expanded but still in virial equilibrium. The general statistical
results of the four models (see Table \ref{tbl-1}) are presented in Table \ref{tbl-2}. More than half of the planetary systems still
retain their original planet number. A cluster can lose about 26\%-34\% of its original born planets. The number of surviving planets
($N_{\rm P}$), FFPs inside the cluster ($N_{\rm FFPi}$), FFPs outside ($r>10$ pc)the cluster ($N_{\rm FFPo}$), and single stars without any
companions ($N_{\rm ss}$) depend on the different bounded energy of planets $E_{\rm b}$.
\item More than 90\% of the 2pisi systems change eccentricities less than 0.1 or inclinations less than 10$^\circ$, while most 1pisi systems have eccentricities or inclinations of planets that obviously changed. Planets in pisj systems have a wide, flat distribution of eccentricities and inclinations. In 1pisi systems, inner planets are
preserved preferentially. If an inner planet was ejected, the remaining planet seems to have more probability of changing eccentricities or
inclinations (Section 4.2).
\item Under the assumption that all the stars spin in a fixed direction in our cluster models, at least 6\% of the stars have misaligned planetary systems and 1\%
have retrograde planets. These spin-orbit misalignment systems are likely to be generated in unstable systems (1pisi or pisj).
\item Unstable planetary systems are concentrated in the inner region of the clusters while the stable systems are following a fatter
distribution in the outer region, as shown in Figure \ref{fig6}(b). With a sharp peak at $\sim$1 pc, the fraction of planetary systems with
$N_{\rm P}=0$ in the inner 2 pc is about 80\%. The fraction of systems with $N_{\rm P}=1,2$ in 2 pc is about 60\% and 50\%
respectively. Our results are constisten with observations:  two of the four planetary sytems are in 2 pc of OCs.
\item The stellar mass is also a key factor for the stability of bounded planets. We obtained a critical mass of $\sim2.5M_\odot$ in Figure 8,
above which the planetary systems are probably unstable and most of them lose at least one planet. Planetary systems around stars with mass
$<2.5M_\odot$ are likely to maintain all their original planets. According to our results, a large fraction ($>80\% $) of the bounded
planets can be found around stars with ($M=0.1$-$1 M_\odot$) in VYOCs. Massive stars ($>16M_\odot$) tend to lose all the planets around them.
We also compared our results to observations.
\item In YOCs, binary systems can be ejected or disrupted in the timescale $\tau_{\rm diss}$ or $\tau_{\rm disrupt}$. The binary fraction can be estimated by Equation (\ref{fb}) in Section 5.1.
In our model, nearly 64\% of the binaries still exist in the cluster after 10 Myr. However their orbits have been changed due to the evolution of the cluster.
The number of binary systems with $a_{\rm b}>200$ AU decreases by disruption, and binary systems with ecc$<0.4$ likely have their eccentricities pumped
to a moderate value. At the same time, a minority of new binary systems (15 in total) have formed.
\item The planets around disrupted binary stars are more unstable than those around single stars. After the violent perturbations during binary disruptions, more than 30\% of the planetary systems lost at least one planet. However, planetary systems around these disrupted binary stars contribute lots of retrograde planets, as pointed out in Section 5.2. 
\item The stability of planetary systems in binaries depends on $a_{\rm b},e_{\rm b}$, as shown in Section 5.2. We give a rough criterion:
planets in binary systems with $a_{\rm b}(1-e_{\rm b}^2)>100$ AU are hardly disrupted during the cluster evolution. The influence of
$r$ on binary systems is less obvious than that in single stars.
\item 15\%-25\% of the planets are released as FFPs, and only 1/4 of them are still cursing in OCs after 10 Myr. However, they can only stay in the inner
domain of the cluster; therefore the CDFs of them correspond to each other. More than 80\% of the FFPs are concentrated in 4 pc, and the
maximum fraction peaks around 1.14 pc, as shown in Figure 13(a) in Section 6.
\item The ejected objects contain $\sim48\%$ fruitful FFPs (see Figure 13(b)). More than 80\% of the ejected stars still have the same initial number of bounded planets. This indicates tht our solar system, which is thought to formed in a cluster environment, is most likely to form as a similar configuration as its current state.
\end{itemize}

However in our cluster models, some assumptions are still too simple. First, the residual gas in the cluster is not included. In
clusters, gas with limited mass can hardly influence the dynamical evolution of the cluster. However, the gas disks around stars play
crucial roles on the formation and evolution of planets \citep{Liu11}. The gravity of the outer gas disk can also lead to secular
effects on multi-planet systems and under some conditions with small mutual inclination can also lead to the onset of the Kozai effect.
\citep{Chen12}. In clusters, the flybys also influence the structure of the gas disk and consequently change the occurrence of planets
\citep{FR09}.

Second, the isotropic assumption and virial equilibrium in our cluster model are also queried by some authors. \citet{SA09} and
\citet{Schmeja11} indicated a substructure in young star-forming regions. As mentioned by \citet{Parker12}, the number of surviving
planets also depends on the initial virial parameter $Q$ of the cluster.

We only choose four models with different planetary systems, therefore it is hard to discuss the influence of bounded energy $E_{\rm
b}$ in detail. In our further work, additional planetary systems are needed to study the correlation between planetary stability
and $E_{\rm b}$. We only set two planets around each star as the first step to consider the multi-planetary systems. The stabilities
of a system with more planets might be much more sensitive with close encounters. Different properties of the clusters, i.e., the total
mass, core radium, number of stars, etc., lead to different timescales of cluster evolution \citep{Mal07b}. Therefore, the fraction of
preserved or ejected planets depends on these parameters too.

The rotation of the cluster will influence the dynamical evolution of cluster directly. Observations of cluster NGC 4244 show an obvious
rotation \citep{Seth08}. Since we adopted a non-rotating cluster, we can only study a one-dimensional $r$-correlation in Section 4.3.
Adding a rotating rate to the clusters, the stability of planetary systems at different latitudes with the same distance is varied due to
the different mean velocities $v$, which is presented in the expression of $\tau_{\rm enc}$ (Equation (\ref{enc})).

As we only study the VYOCs here, the galactic tidal effect and stellar evolution are not important, as pointed out in Section 2.1. However, when studying
a longer evolution of the cluster with age $>10$ Myr, the galactic tides and stellar evolution timescales need be estimated
again. Galactic tides tend to evaporate cluster members and change the properties of clusters \citep{Baum03}. During stellar evolution,
a star will experience red giant branch, horizontal branch, asymptotic giant branch, etc., in the H-R diagram. The stability of planets around it must be checked carefully
during all these phases \citep{VL07}.

%

%




{\bf Acknowledgements} This work is supported by the National Basic Research Program of China (2013CB834900), National Natural Science
Foundations of China (Nos. 10833001, 10925313, and 11078001), National Natural Science Funds for Young Scholar (No. 11003010), Fundamental
Research Funds for the Central Universities (No. 1112020102), and the Research Fund for the Doctoral Program of Higher Education of
China (Nos. 20090091110002 and 20090091120025).

\clearpage

\begin{table}
\caption{\textnormal{Masses and Locations of Two Planets in Different Models with Initial
Eccentricity=0,Inclination=1$^\circ$.}\label{tbl-1}}.
\begin{center}
\begin{tabular}{lcccc}
\hline
Model & $M_{\rm p1}$  & $a_{\rm p1}$(AU)  & $M_{\rm p1}$ & $a_{\rm p1}$(AU)\\
\hline
2J     & 1 $M_{\rm J}$ & 5.2 & 1 $M_{\rm J}$ & 9.5  \\
J10E4  & 1 $M_{\rm E}$ & 4   & 1 $M_{\rm J}$ & 10.4\\
J10E2  & 1 $M_{\rm E}$ & 2   & 1 $M_{\rm J}$ & 10.4\\
J5E2   & 1 $M_{\rm E}$ & 2   & 1 $M_{\rm J}$ & 5.2  \\
\hline
\end{tabular}
\end{center}
\end{table}

\begin{table}
\caption{\textnormal{General Results of Different Four Models at $t=10$ Myr.}}\label{tbl-2}.
\begin{center}
\begin{tabular}{lccccccccccc}
\hline
Model & $N_{\rm P}$  & $N_{\rm S}$  & $N_{\rm 2pisi}$ & $N_{\rm 1pisi}$ & $N_{\rm pisj}$ & $N_{\rm 0ps}$ & $N_{\rm FFPi}$ & $N_{\rm FFPo}$ & $N_{\rm b}$  & $N_{\rm ss}$ & $E_{\rm b}$ \\
\hline
2J    & 1423 & 852 & 582 & 154 & 9 & 107 & 96 & 394 & 160   & 19 & ...\\
J10E4 & 1320 & 799 & 553 & 115 & 5 & 126 & 94 & 416 & 153   & 30 & Low\\
J10E2 & 1479 & 853 & 610 & 176 & 4 & 63  & 79 & 327 & 154   & 19 & Moderate\\
J5E2  & 1486 & 810 & 685 & 60  & 2 & 63  & 54 & 244 & 137   & 14 & High\\
\hline\\
\end{tabular}
\end{center}
{\bf Notes.} 66\%-74\% of planets and 80\%-85\% of stars still
remain in the clusters. More than half of planetary systems still have two original planets, while 6\%-18\% lost one planet. The
recaptured planets are very rare, $<1\%$. Besides these bounded planets, 2.7\%-5\% of planets became FFPs cruising in clusters,
while 12\%-21\% became FFPs outside the host clusters. $\ast {\rm N}_{\rm P}$ and $N_{\rm S}$ represent the number of surviving planets and stars in the cluster. $N_{\rm
2pisi}$, $N_{\rm 1pisi}$, $N_{\rm pisj}$, and $N_{\rm 0ps}$ represent the number of systems with two original planets, only one
original planet, only one recaptured planet and no planet, respectively. $N_{\rm FFPi}$ is the number of FFPs staying in clusters,
while $N_{\rm FFPo}$ is FFPs ejected outside, and $N_{\rm b}$ is the number of binary pairs. The number of single stars with no planets
or stellar companions is $N_{\rm ss}$. The bounded energy $E_{\rm b}$ is the total initial energy of two planets around the
host star.
\end{table}

\begin{figure}
\vspace{0cm}\hspace{0cm} \epsscale{1} \plotone{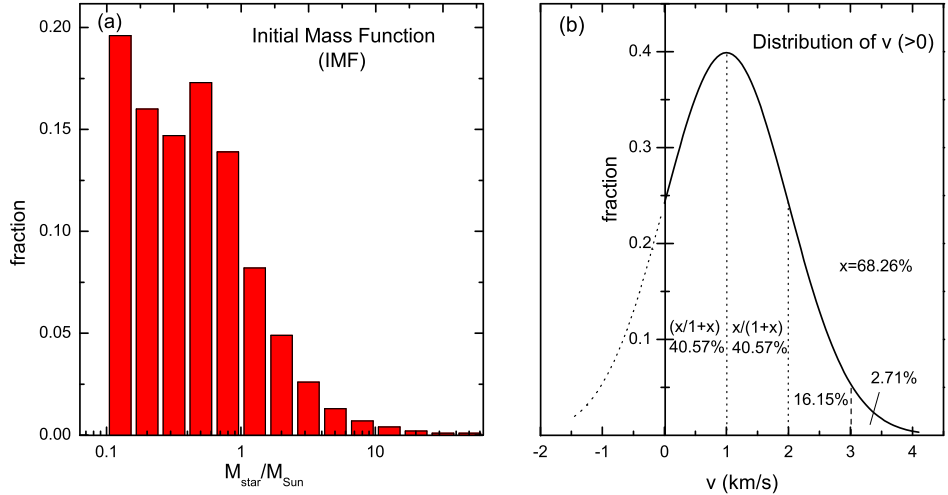} \vspace{0cm} \caption{(a) The initial mass function (IMF) of our cluster
model. (b) The Gaussian distribution of stellar velocities; we truncated the distribution when ($v<0$ km s$^{-1}$). Only about 3\% of the stars
have velocities $>3$ km s$^{-1}$. \label{fig1}}
\end{figure}

\begin{figure}
\vspace{0cm}\hspace{0cm} \epsscale{1} \plotone{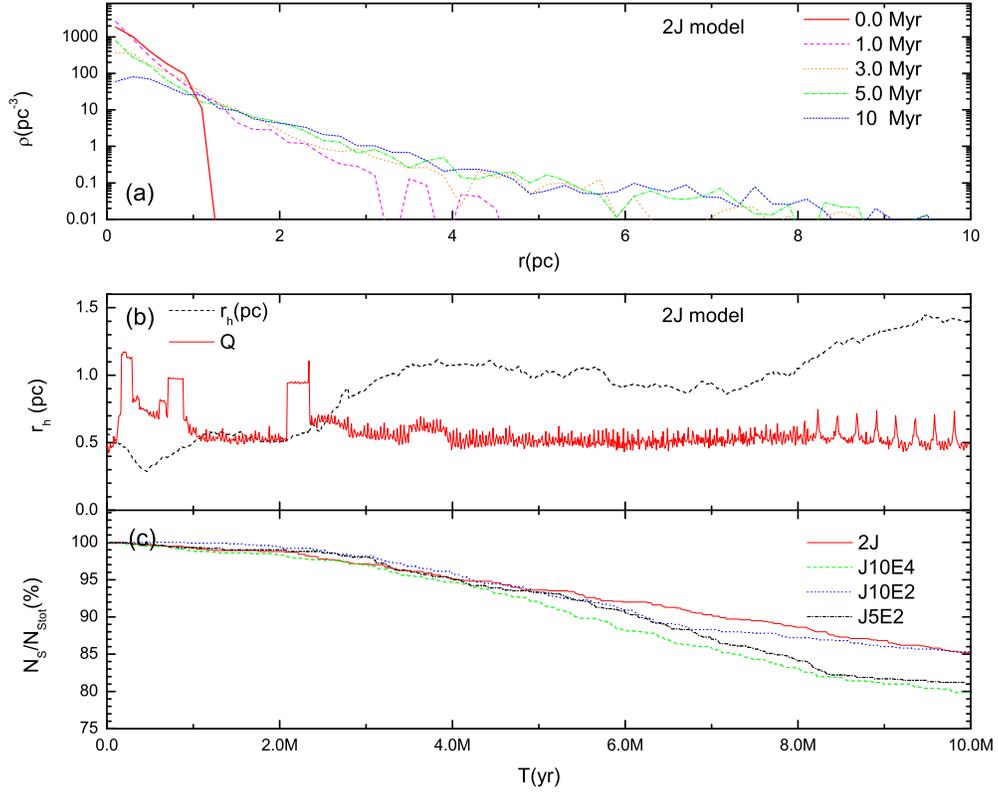} \vspace{0cm} \caption{(a) The stellar densities in the 2J model at $t$=0, 1, 3, 5,
10 Myr. The stellar density sharply decreases with time. After 10 Myr, the density is at least one order smaller than the initial
values. (b) The variations of virial parameter $Q$ (red line) and half mass radius $r_{\rm h}$ (black line) with time. $Q$ can keep
0.5 after 3 Myr which means the virial equilibrium is nearly satisfied. As $r_h$ increases, the cluster expands quickly and
$r_h\simeq1.5$ pc after 10 Myr, about three times larger than the initial value. (c) The fraction of stars in clusters decreases with time
in different models. In the first initial 3 Myr, the fraction decreases slowly ($\sim$3\%), after that the fraction decreases to
80\%-85\%. \label{fig2}}
\end{figure}

\begin{figure}
\vspace{0cm}\hspace{0cm} \epsscale{1} \plotone{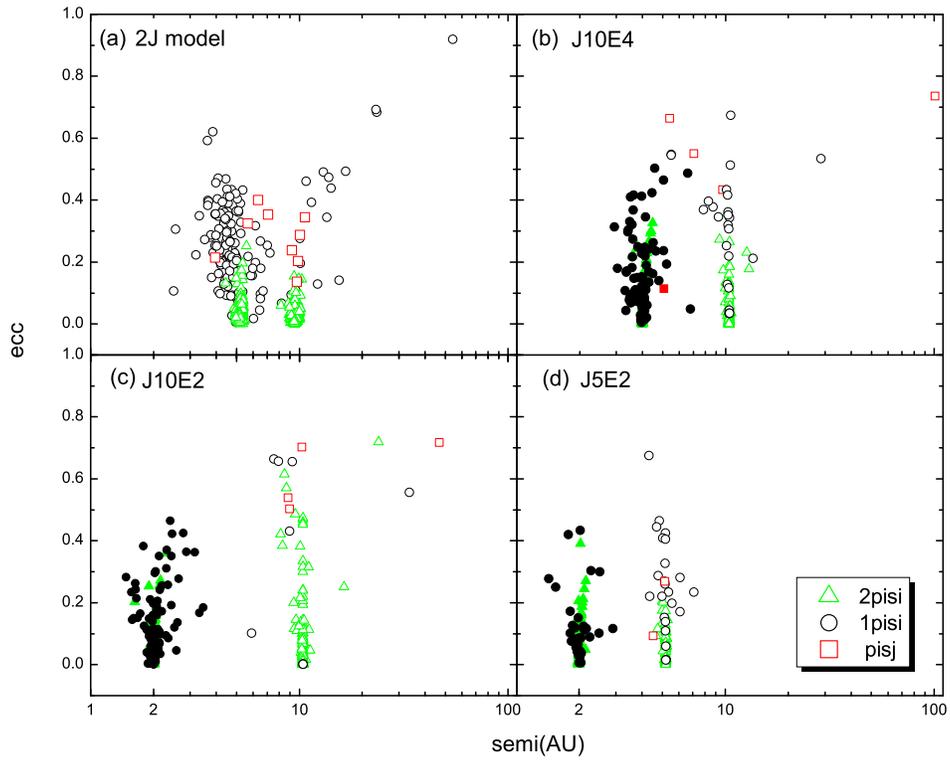} \vspace{0cm} \caption{Configurations of survival planetary systems for four
different models. Three classes of systems are represented here: 2pisi (green triangles), 1pisi (black circles), and pisj (red squares)
systems. The filled symbols represent Earth-like planets while the open symbols represent Jupiter-like planets. Earth-like planets are hard to
eject therefore most 1pisi systems are shown as filled circles. \label{fig3}}
\end{figure}

\begin{figure}
\vspace{0cm}\hspace{0cm} \epsscale{1} \plotone{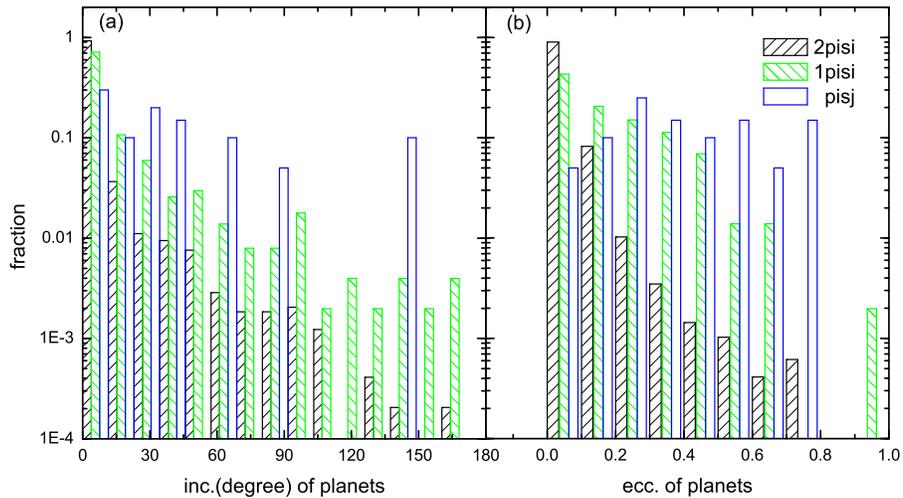} \vspace{0cm} \caption{Distributions of (a) inclinations  and
(b) eccentricities of 2pisi, 1pisi, and pisj systems. Most 2pisi systems have little change in their eccentrics and inclinations, while most 1pisi
systems have eccentric or inclined planets. Recaptured planets tend to stay in very eccentric and inclined orbits. \label{fig4}}
\end{figure}

\begin{figure}
\vspace{0cm}\hspace{0cm} \epsscale{1} \plotone{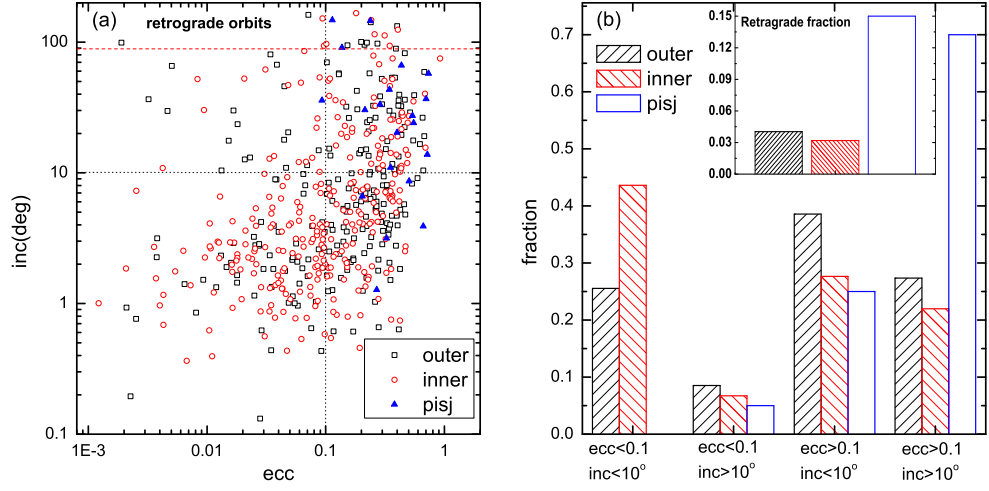} \vspace{0cm} \caption{(a) Distributions of eccentricities (ecc)  and
inclinations (inc) in single planetary systems. The ecc-inc panel is divided into four regions by two black dotted lines: ecc = 0.1 and
inc = 10$^{\circ}$. The red line corresponds to inc = 90$^\circ$. Planets above this line move in retrograde orbits. (b) Planetary
fractions in four different regions. In 1pisi systems, the inner planet (red circles) is more reserved than the outer
planet (black squares). As the outer one is ejected, about 55\% of planetary systems have changed ecc or inc obviously. However if the
inner one is ejected, more than 70\% of planetary systems have obviously changed ecc or inc. No recaptured planets have small
eccentricities and inclinations. \label{fig5}}
\end{figure}

\begin{figure}
\vspace{0cm}\hspace{0cm} \epsscale{1} \plotone{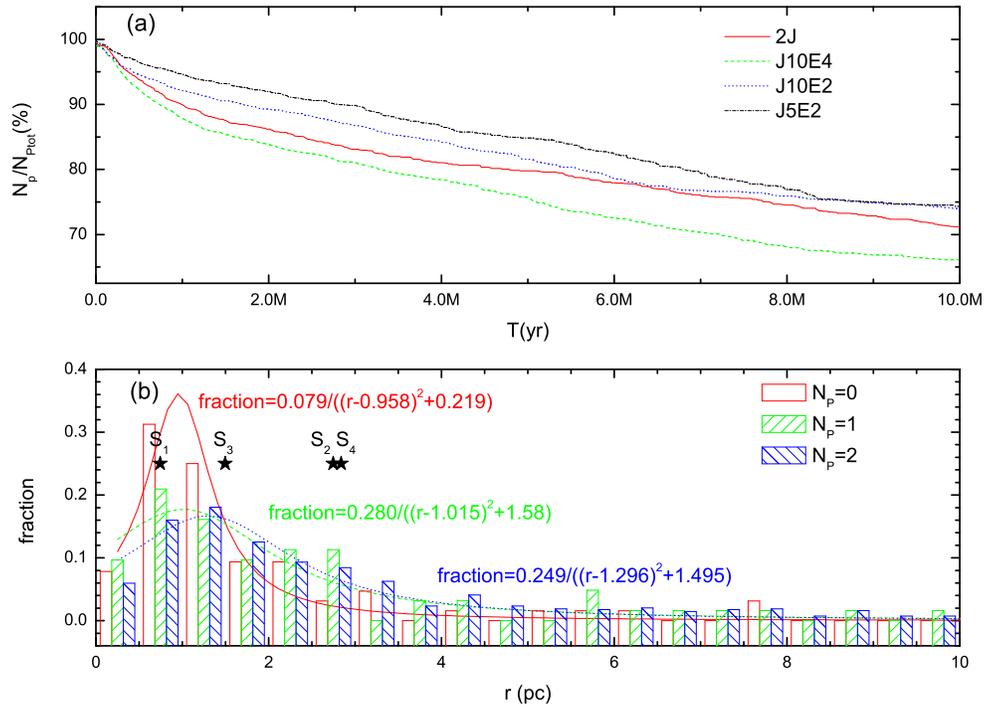} \vspace{0cm} \caption{(a) The fraction of planets in the cluster decreases
with time in the four different models. In model J5E2, which has a larger bounded energy, more planets left, while in model J10E4, on the contrary, fewer planets left. (b) The fraction of planetary systems with $N_{\rm P}= 2,1,0$ (the number of surviving planets, hereafter the
same) are still related to $r$ (the distance from the cluster center). The distribution of unstable planetary systems with $N_{\rm
P}=0$ peaks at 0.958 pc sharply, while a fatter distribution of systems with $N_{\rm P}=1$ peaks around 1.015 pc. The peak for stable
systems with $N_{\rm P}=2$ is located at 1.29 pc, i.e., these stable systems stay in the outer region compared with the other two
unstable systems. Label $S_1,S_2,S_3$, and $S_4$ represent the host stars of the four known planets in OCs, i.e., No. 3 in NGC 2423, $\epsilon$
Tauri in Hyades, Pr0201 and Pr0211 in Praesepe, respectively. Their locations are from the WEBDA database
(http://www.univie.ac.at/webda).} \label{fig6}
\end{figure}

\begin{figure}
\vspace{0cm}\hspace{0cm} \epsscale{1} \plotone{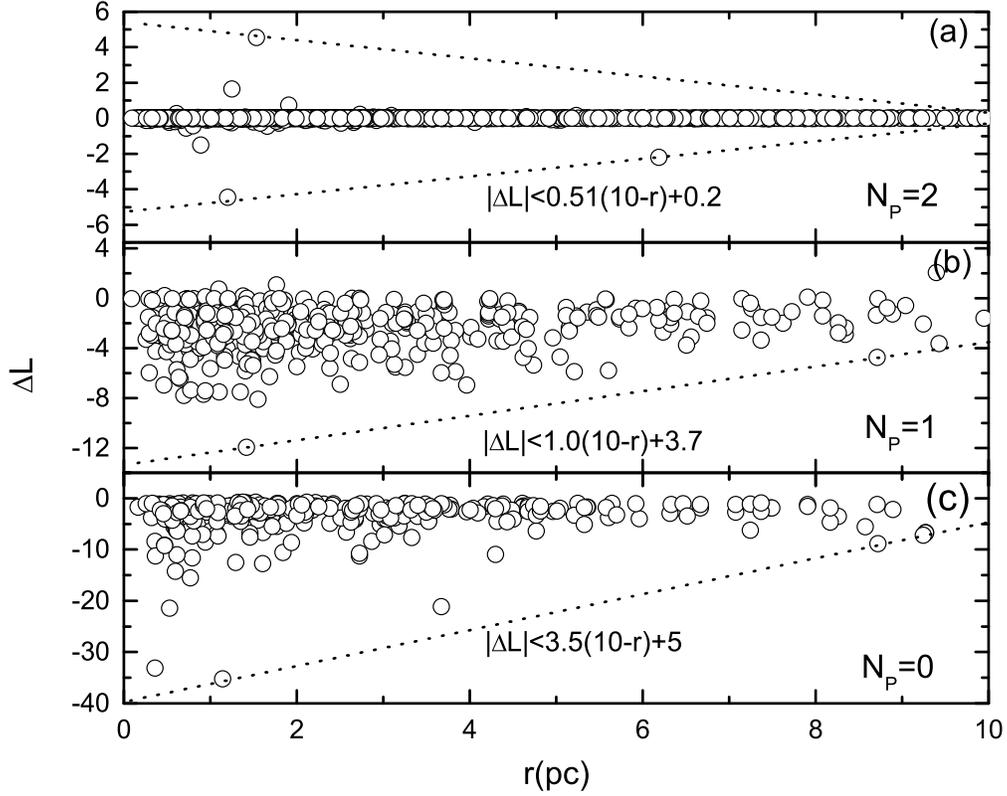} \vspace{0cm} \caption{Variations of total angular momentum $\Delta L$ for
planets at different cluster locations (dist) for $N_{\rm P}=2,1,0$ from top to bottom. The unit of $\Delta L$ is $M_{\rm
J}\cdot\sqrt{GM_\odot\cdot {\rm AU}}$). We can find the correlation between $\Delta L$ and $r$, shown by dotted lines: $|\Delta
L|<a(10-r)+b$. The constants $a,b$ depend on $N_{\rm P}$. Much denser environments make the angular momentum exchanges more
frequently and lead to a larger $\Delta L$. \label{fig7}}
\end{figure}

\begin{figure}
\vspace{0cm}\hspace{0cm} \epsscale{1} \plotone{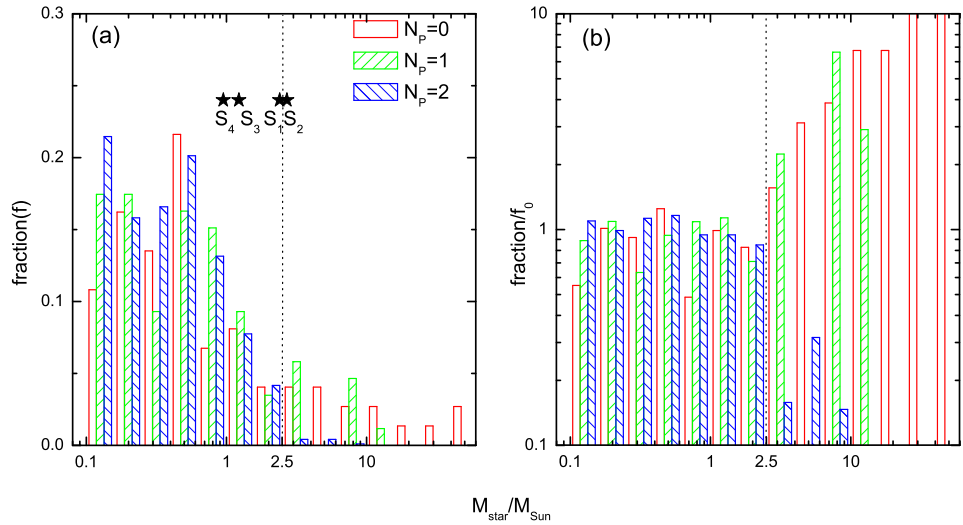} \vspace{0cm} \caption{(a) The fraction of planetary systems ($N_{\rm
P}=0,1,2$) correlates with stellar masses. Observational data with labels $S_1,S_2,S_3$, and $S_4$ are the same as those in
Figure \ref{fig6}(b), and 3($75\%$) of them are less than $2.5 M_\odot$ (the dotted line). (b) The normalized fraction correlates with stellar
masses. As shown in panel (b), if the host star has a mass $<2.5M_\odot$, the normalized fraction is around unit, i.e., it is nearly
the same fraction as the IMF. However, a star more massive than $2.5M_\odot$ tends to lose at least one planet; therefore the normalized
fractions of 1pisi and pisj systems are obvously enhanced.\label{fig8}}
\end{figure}

\begin{figure}
\vspace{0cm}\hspace{0cm} \epsscale{1} \plotone{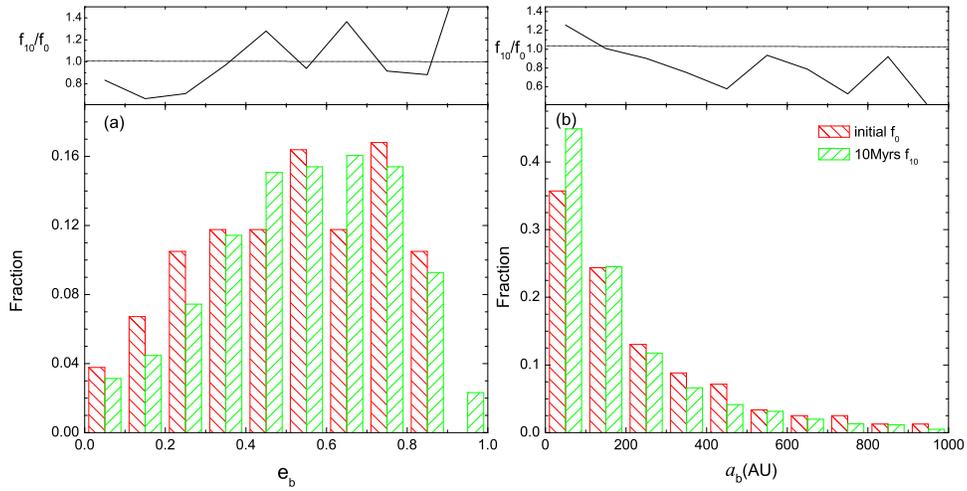} \vspace{0cm} \caption{Orbital properties of binary systems in a cluster
after 10 Myr: the distribution of (a) eccentricities $e_{\rm b}$ and (b) semi major axes $a_{\rm b}$ of binary systems. During the
dynamical evolution of a cluster, the gravity of stars can disrupt wide binary systems or pump the eccentricities of binary systems.
Therefore, compared with the initial distribution, the fraction of binary systems with small $e_{\rm b}(<0.4)$ or large $a_{\rm
b}(>200 $AU) decreases. \label{fig9}}
\end{figure}

\begin{figure}
\vspace{0cm}\hspace{0cm} \epsscale{1} \plotone{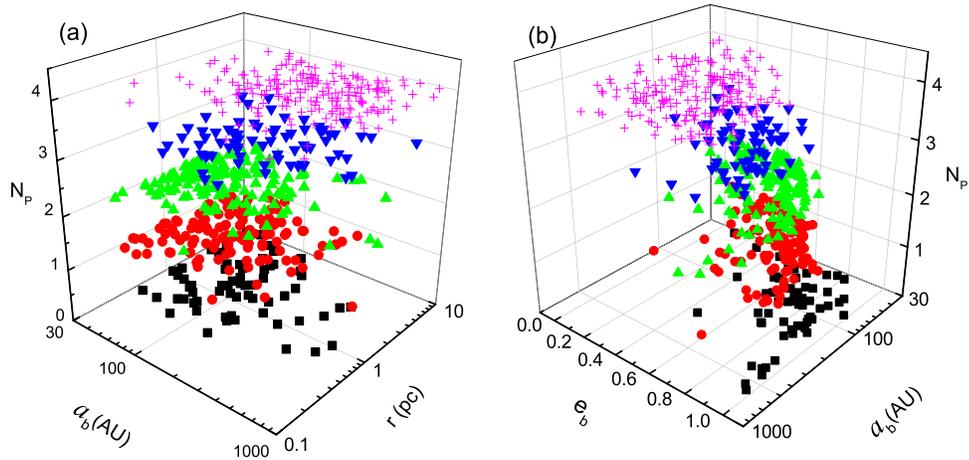} \vspace{0cm} \caption{Number of surviving planets in binary systems
$N_{\rm P}$ in the (a) $a_{\rm b}$-$r$ and (b) $e_{\rm b}$-$a_{\rm b}$ planes. $a_{\rm b},e_{\rm b}$ are the semi major axis and
eccentricity of a binary system, and $r$ is the distance from the barycenter to the center of the cluster. The pink crosses, blue inverted
triangles, green triangles, red circles, and black squares represent $N_{\rm P}=4, 3, 2, 1, 0$, respectively. As shown in panels (a) and
(b), the binary systems with larger angular momentum (larger $a_{\rm b}$ and smaller $e_{\rm b}$) and smaller stellar density (larger
$r$) can probably hold on to more planets. \label{fig10}}
\end{figure}

\begin{figure}
\vspace{0cm}\hspace{0cm} \epsscale{1} \plotone{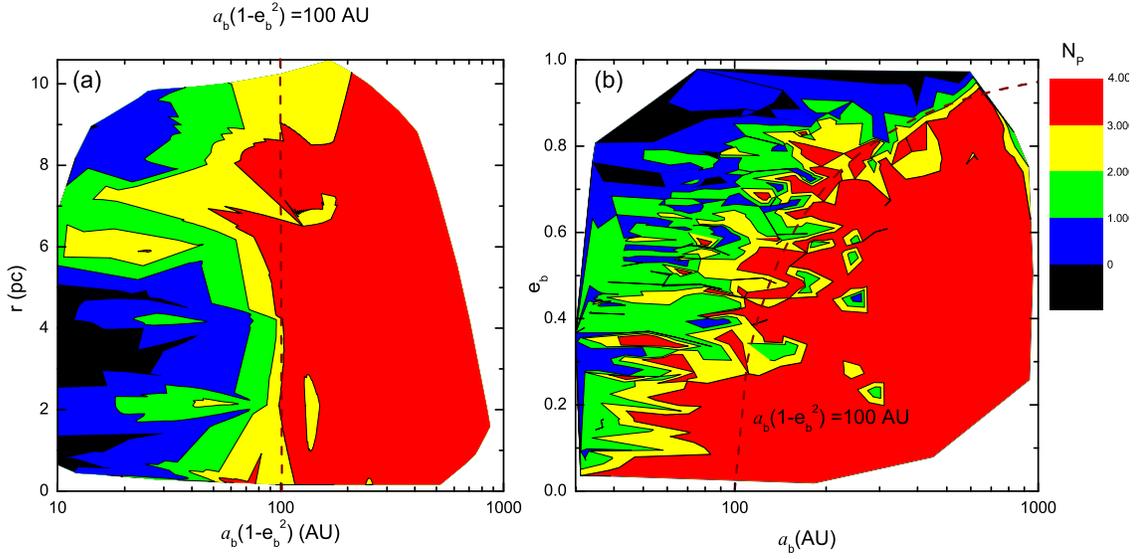} \vspace{0cm} \caption{(a) The number of surviving planets $N_{\rm P}$ in
binary systems in the $X$-$r$ plane ($X$=$a_{\rm b}(1-e_{\rm b}^2)$),  $r$ is the distance from the center of the star cluster; (b) $N_{\rm P}$ in the
$a_{\rm b}$-$e_{\rm b}$ plane. There is an obvious boundary: planets in binary systems with $a_{\rm b}(1-e_{\rm b}^2)>100$ AU are
hardly disrupted during the evolution of the cluster. \label{fig11}}
\end{figure}

\begin{figure}
\vspace{0cm}\hspace{0cm} \epsscale{1} \plotone{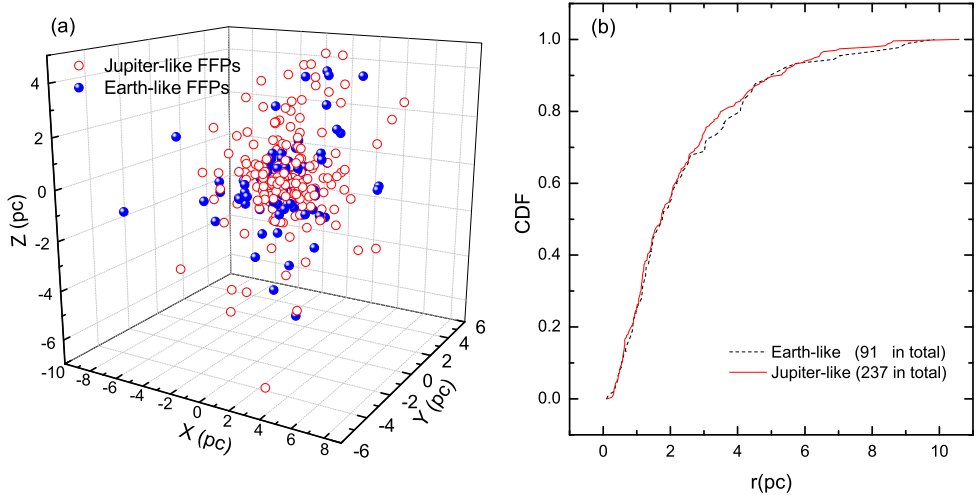} \vspace{0cm} \caption{ (a) The spatial distributions of Jupiter-like (red
open circles) and Earth-like (blue filled circles) FFPs; (b) the CDFs of Jupiter-like(red solid line) and Earth-like(blue dash line)
FFPs. There are no differences between their CDFs. There are about 2.5 times more Jupiter-like FFPs  than Earth-like planets. \label{fig12}}
\end{figure}

\begin{figure}
\vspace{0cm}\hspace{0cm} \epsscale{1} \plotone{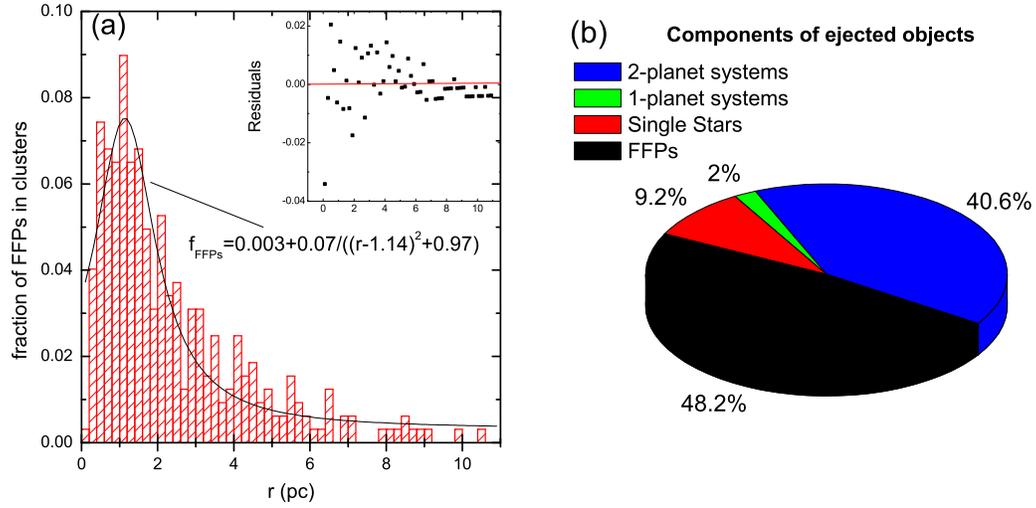} \vspace{0cm} \caption{(a) The distribution of FFPs in clusters; about one
half of FFPs are concentrated in the center of the clusters ($< 2$ pc), and the maximum fraction is around 1.14 pc. (b) Different
components of ejected objects; nearly half($\sim48.2\%$) of them are FFPs. About $40.6\%$ of ejected objects are still original
two-planet systems, only $9.2\%$ of stars have no companions. one-planet systems(either 1pisi or pisj systems) are very rare. Eighty percent of
these ejected stars still have two original planets around them. \label{fig13}}
\end{figure}


\begin{thebibliography}{99}
\bibitem[{Adams et al.(2006)}]{Adams06} Adams, F. C., Proszkow, E. M., Fatuzzo, M. \& Myers, P. C., 2006, ApJ, 641, 504
\bibitem[{Adams(2010)}]{Adams10} Adams, F. C., 2010, ARA\&A, 48, 47
\bibitem[{Baumgardt \& Makino,(2003)}]{Baum03}Baumgardt, H. \& Makino, J., 2003, MNRAS, 340, 227
\bibitem[Bihain et al.(2009)]{Bihain09}Bihain, G., Rebolo, R., Zapatero, O. M. R., et al. 2009, A\&A, 506, 1169
\bibitem[Binney \& Tremaine(1987)]{Binn87}Binney, J., \& Tremaine, S., 1987, Galactic Dynamics (Princeton, NJ: Princeton Univ. Press), 747
\bibitem[{Bonatto \& Bica,(2009)}]{Bonatto09}Bonatto, C., \& Bica, E. 2009, MNRAS, 394, 2127B
\bibitem[{Bonatto et al.,(2006)}]{Bonatto06}Bonatto, C., Santos, J. F. C., Jr., \& Bica, E. 2006, A\&A, 445,567
\bibitem[{Bressan et al.,(1993)}]{Bressan93}Bressan, A., Fagotto, F., Bertelli, G. \& Chiosi, C., 1993, A\&AS, 100, 647

\bibitem[Chambers(1999)]{Cham99}Chambers, J. E., 1999, MNRAS, 304, 793
\bibitem[Chen et al.(2012)]{Chen12} Chen, Y.-Y., Liu, H.-G. \& Zhou, J.-L., 2012, ApJ. 769, 26

\bibitem[Davies \& Sigurdsson(2001)]{DS01}Davies, M. B., \& Sigurdsson, S. 2001, MNRAS, 324, 612
\bibitem[Duquennoy \& Mayor(1991)]{DM91}Duquennoy, A. \& Mayor, M., 1991, A \& A, 248, 485
\bibitem[Fischer \& Marcy(1992)]{FM92}Fischer, D. A. \& Marcy, G. W., 1992, ApJ, 396, 178
\bibitem[Forgan \& Rice(2009)]{FR09} Forgan, D. \& Rice, K., 2009, MNRAS, 400, 2022
\bibitem[Gieles \& Baumgardt(2008)]{GB08}Gieles, M. \& Baumgardt, H., 2008, MNRAS, 389, 28
\bibitem[{Hern\'{a}ndez et al.,(2007)}]{Hernandez07}Hern\'{a}ndez, J., Hartmann, L., Megeath, T., et al., 2007, ApJ, 662, 1067
\bibitem[Holman \& Wiegert (1999)]{MP99}Holman, M. J. \& Wiegert, P. A., 1999, AJ, 117, 621

\bibitem[King,(1966a)]{King66}King, I. R. 1966, AJ, 71, 64
\bibitem[{Kozai(1962)}]{Koz62} Kozai, Y., 1962, \aj, 67, 591
\bibitem[Kraus, et al.(2007)]{Krause07}Kraus, S., Balega, Y. Y., Berger, J.-P., et al., 2007, A \& A, 466, 649
\bibitem[Kraus, et al.(2009)]{Krause09}Kraus, S., Weigelt, G., Balega, Y. Y., et al., 2009, A \& A, 497, 195
\bibitem[Kroupa(2002)]{Kroupa02}Kroupa, P., 2002, Sci, 295, 82
\bibitem[Kroupa,(2008)]{Kroupa08}Kroupa, P., 2008, in initial Conditions for Star Clusters, ed. S. J. Aarseth, C. A. Tout, \& R. A. Mardling (Lecture Notes in Physics, Vol. 760; Berlin: Springer), 181

\bibitem[{Lada \& Lada(2003)}]{Lada03}Lada, C. J. \& Lada, E. A., 2003, ARA\&A, 41, 57
\bibitem[{Lada(2010)}]{Lada10}Lada, C. J., 2010, RSPTA, 368, 713
\bibitem[Laughlin \& Adams(1998)]{LA98}Laughlin, G., \& Adams, F. C. 1998, ApJL, 508, L171
\bibitem[{Liu et al.,(2011)}]{Liu11} Liu, H. G.,  Zhou,J.L., \& Wang, S., 2011, ApJ, 732, 66
\bibitem[{Lovis \& Mayor(2007)}]{Lovis07}Lovis, C. \& Mayor, M., 2007, A\&A, 472, 657
\bibitem[{Lucas \& Roche(2000)}]{Lucas00}Lucas, P. W., \& Roche, P. F., 2000, MNRAS, 314, 858

\bibitem[{Malmberg et al.(2007a)}]{Mal07a} Malmberg, D., Davies, M.~B., \& Chambers, J.~E., 2007a, \mnras, 377, L1
\bibitem[{Malmberg et al.(2007b)}]{Mal07b} Malmberg, D., de Angeli, F., Davies, M.~B., et al., 2007b, MNRAS, 378, 1207
\bibitem[{Malmberg et al.(2011)}]{Mal11} Malmberg, D., Davies, M.~B., \& Heggie, D.~C., 2011, MNRAS, 411, 859
\bibitem[Mason, et al.(2009)]{Mason09}Mason, B. D., Hartkopf, W. I., Gies, D. R., Henry, T. J. \& Helsel, J. W., 2009, AJ, 137, 3358
\bibitem[Holman \& Wiegert (1999)]{MP99}Holman, M. J. \& Wiegert, P. A., 1999, AJ, 117, 621
\bibitem[Mayor et al.(1992)]{Mayor92}Mayor, M., Duquennoy, A., Halbwachs, J.-L. \& Mermilliod, J.-C., 1992, in ASP Conf. Ser. 32, IAU Colloq. 135, Complementary Approaches to Double and Multiple Star Research, ed. H. A. McAlister \& W. I. Hartkopf (San Francisco, CA: ASP), 73
\bibitem[{Nagasawa \& Ida,(2011)}]{Nag11}Nagasawa, M. \& Ida, S., 2011, ApJ, 742, 72N

\bibitem[{Parker \& Goodwin(2009)}]{PG09}Parker, R. J., \& Goodwin, S. P., 2009, MNRAS, 397, 1041
\bibitem[{Parker \& Quanz(2012)}]{Parker12}Parker, R. J., \& Quanz, S. P., 2012, MNRAS, 419, 2448P
\bibitem[Quinn et al.(2012)]{Quinn12}Quinn, S. N., White, R. J., Latham, D. W., et al. 2012, ApJ, 756, 33
\bibitem[Raghavan et al.(2010)]{Rag10}Raghavan, D., McMaster, H. A., Henry, T. J. et al., 2010, ApJSS, 190, 1

\bibitem[{S\'{a}nchez \& Alfaro(2009)}]{SA09}S\'{a}nchez, N., \& Alfaro, E. J., 2009, ApJ, 696, 2086
\bibitem[{Sato et al.(2007)}]{Sato07}Sato, B., Izumiura, H., Toyota, E., et al., 2007, ApJ, 661, 527
\bibitem[{Schmeja(2011)}]{Schmeja11}Schmeja, S., 2011, AN, 332, 172
\bibitem[{Seth, et. al.(2008)}]{Seth08}Seth, A. C., Blum, R. D., Bastian, N. et al., 2008, ApJ, 687, 997
\bibitem[Smith \& Bonnell(2001)]{SB01}Smith, K. W., \& Bonnell, I. A. 2001, MNRAS, 322, L1
\bibitem[Spitzer(1987)]{Spitzer87}Spitzer, L. 1987, Dynamical Evolution of Globular Clusters. (Princeton, NJ: Princeton Univ. Press), 40
\bibitem[{Spurzem et al.(2009)}]{Spurzem09}Spurzem, R., Giersz, M., Heggie, D. C., \& Lin, D. N. C., 2009, ApJ, 697, 458
\bibitem[{Sumi et al.,(2011)}]{Sumi11}Sumi, T., Kamiya, K., Bennett, D. P., et al., 2011, Natur, 473, 349

\bibitem[Terquem \& Papaloizou(2002)]{TP02}Terquem, C., \& Papaloizou, J. C. B. 2002, MNRAS, 332, L39
\bibitem[Villaver \& Livio(2007)]{VL07}Villaver, E., \& Livio, M., 2007, ApJ, 661, 1192
\bibitem[Villaver \& Livio(2009)]{VL09}Villaver, E., \& Livio, M. 2009, ApJL, 705, L81
\bibitem[{Wang et al.(2011)}]{Wang11}Wang, J., Feigelson, E. D., Townsley, L. K., et al., 2011, ApJS, 194, 11W
\bibitem[{Winn et al.(2010)}]{Winn10}Winn, J. N., Fabrycky, D., Albrecht, S. \& Johnson, J. A., 2010, ApJL, 718, L145
\bibitem[{Wu \& Lithwick,(2011)}]{Wu11}Wu, Y., \& Lithwick, Y., 2011, ApJ, 735, 109
\bibitem[Zhou et al.(2007)]{Zhou07}Zhou, J. L., Lin, D. N. C., \& Sun, Y.-S. 2007, ApJ, 666, 423
\bibitem[Zhou et al.(2012)]{Zhou12} Zhou, J.L., Xie, J.W., Liu,  H.G.,  Zhang,H., \& Sun,Y.S. 2012 RAA, 12, 1081
\end{thebibliography}
\end{document}